\documentclass[aps,prb,letterpaper,amsmath,amssymb,reprint,superscriptaddress]{revtex4-1}
\usepackage{graphicx}

\begin{document}
\title{Insights into the electronic structure of Co$_2$FeSi from x-ray magnetic linear dichroism}
\author{Markus Meinert}
\email{meinert@physik.uni-bielefeld.de}
\author{Jan-Michael Schmalhorst}
\author{Manuel Glas}
\author{G\"unter Reiss}
\affiliation{Thin Films and Physics of Nanostructures, Department of Physics, Bielefeld University, D-33501 Bielefeld, Germany}
\author{Elke Arenholz}
\affiliation{Advanced Light Source, Lawrence Berkeley National Laboratory, CA 94720, USA}
\author{Tim B\"ohnert}
\author{Kornelius Nielsch}
\affiliation{Institute of Applied Physics, University of Hamburg, Jungiusstrasse 11, D-20355 Hamburg, Germany}

\date{\today}

\begin{abstract}
Experimental evidence both for and against a half-metallic ground-state of the Heusler compound Co$_2$FeSi has been published. Density functional theory based calculations suggest a non half-metallic ground state. It has been argued, that on-site Coulomb interaction of the \textit{d} electrons has to be taken into account via the LDA$+U$ method, which predicts a half-metallic ground-state for $U\approx 2.5\dots4.5$\,eV. X-ray magnetic linear dichroism (XMLD) can be used as a tool to assess the appropriateness of the LDA$+U$ approach: the calculated spectra within the LDA$+U$ or GGA$+U$ schemes are different from those within the LDA or GGA. Due to its ability to separate different orbital symmetries, XMLD allows us to distinguish between different models of the electronic structure of Co$_2$FeSi. In this article we discuss experimental XMLD spectra and compare them with detailed first principles calculations. Our findings give evidence for the inadequacy of the LDA$+U$ or GGA$+U$ band structures, whereas constrained calculations with the GGA and a fixed spin moment of 6\,$\mu_\mathrm{B}$ give better overall agreement between experiment and theory.
\end{abstract}

\maketitle

\section{Introduction}
Half-metallic Heusler compounds of the type Co$_2$YZ (with Y a 3\textit{d} transition metal and Z an \textit{sp} main group element) are promising materials for spin electronics. They were shown to follow the Slater-Pauling rule, which gives their magnetization per formula unit as $m = N_v - 24$, where $N_v$ is the number of valence electrons.\cite{Galanakis02} Co$_2$FeSi is the Heusler compound with the highest observed magnetization of nearly $6\,\mu_\mathrm{B}$ per formula unit.\cite{Wurmehl05}

The electronic structures of half-metallic Heusler compounds are routinely computed within the framework of density functional theory (DFT). The local (spin) density and generalized gradient approximations (LDA or GGA) are often able to reproduce the observed ground-state magnetic moments in agreement with experimental data. However, Co$_2$FeSi is a case, for which these approximations fail: the calculated magnetic moment is too low and the density of states is not half-metallic. Because of this, approaches to go beyond the (semi-)local approximations were tested. Applying LDA$+U$ in the fully localized limit with U around 4\,eV opens a minority gap around $E_\mathrm{F}$ and consequently gives the full magnetic moment.\cite{Wurmehl05} Calculations with the non-empirical hybrid PBE0 functional resulted in a half-metallic ground state with a 1\,eV wide minority gap.\cite{Nourmohammadi10} The LDA$+U$ (or GGA$+U$) approach has been applied to study the electronic structure of Co$_2$FeSi and related compounds.\cite{Balke06, Kandpal07, Kumar09, Kim11, Feng12}

Although Co$_2$FeSi satisfies the Slater-Pauling rule and has a near-integer magnetic moment per formula unit, which is a necessary condition for half-metallic ferromagnetism, there is some experimental evidence to the contrary. Tunnel magnetoresistance experiments show rather low TMR with Co$_2$FeSi electrodes, and the conductance curves lack characteristic features of the minority gap.\cite{Lim08,Kubota09} Further, the Gilbert damping is quite high.\cite{Kubota09} On the other hand, a temperature independent resistivity is observed for bulk single- and polycrystalline Co$_2$FeSi below 50\,K, which is seen as a fingerprint of half-metallicity.\cite{Blum09}

Recently, Kallmayer \textit{et al.} predicted from first-principles calculations the x-ray magnetic linear dichroism (XMLD) of Co$_2$FeSi within the LDA and the LDA$+U$ schemes.\cite{Kallmayer11} No significant differences of the x-ray absorption or circular dichroism spectra were predicted. However, clearly different XMLD spectra were computed, and XMLD was proposed as a tool to determine which electronic structure is actually present in Co$_2$FeSi. In this article we use the proposed method on epitaxial thin films of Co$_2$FeSi.

The article is organized as follows: Sections \ref{Sec:Experiment} and \ref{Sec:Computational} describe the experimental conditions and details of our computational approach. The results of the x-ray magnetic circular (XMCD) and linear dichroism measurements and calculations are presented in Section \ref{Sec:Results}, followed by a discussion of the results in Section \ref{Sec:Discussion}. The main part of this article is focused on the understanding of the XMLD of Co$_2$FeSi.  To make sure that the dichroic signals are intrinsical to Co$_2$FeSi, we investigated samples prepared with different deposition conditions.

\section{Experiment}\label{Sec:Experiment}

\begin{table*}[t]
\caption{Preparation conditions of the samples.}
\begin{ruledtabular}
\begin{tabular}{ l l l}
Name & Stack (thicknesses in nm) \\ \hline
 CFS-Cr & MgO(001) / Cr 5 / 700$^\circ$C \textit{in situ} / CFS 20 @ RT/ Mg 0.5 / MgO 1.0 / 450$^\circ$C \textit{ex situ} \\
 CFS-HTD & MgO(001) / CFS 20 @ 700$^\circ$C / Mg 0.5 / MgO 1.0 \\
\end{tabular}
\end{ruledtabular}
\label{Tab:conditions}
\end{table*}

Epitaxial thin films of Co$_2$FeSi have been grown on MgO(001) substrates by dc and rf magnetron co-sputtering with and without seed layers and various heat treatments.

The samples were deposited with a co-sputtering tool with 3" sputter sources. It allows heating of the substrates to 900$^\circ$C. The base pressure was typically $2 \cdot 10^{-9}$\,mbar, and the Ar working pressure was $2 \cdot 10^{-3}$\,mbar. The target-to-substrate distance was 21\,cm. Elemental Co, Fe, and Si targets of 99.9\,\% purity were used. The metals were dc sputtered and Si was rf sputtered. The MgO capping layers were deposited by electron beam evaporation.

We distinguish two types of samples, which are explained in Table \ref{Tab:conditions}: CFS-Cr (with Cr seed layer and post-annealing) and CFS-HTD (high-temperature deposition without seed layer).

X-ray absorption (XAS), x-ray magnetic circular (XMCD) and linear dichroism (XMLD) spectra were taken at beamline 4.0.2 of the Advanced Light Source, Berkeley. Magnetic fields of 0.55\,T were applied to magnetically saturate the samples. XMCD spectra were taken with an angle of 30$^\circ$ between the beam and the sample surface, the magnetic field was parallel to the beam. The circular polarization degree was 90\%. XMLD spectra were taken in normal incidence, with the magnetic field canted out of the sample plane by 10$^\circ$. The degree of linear polarization was 100\%. The magnetization was switched at every photon energy to obtain the dichroic spectra. All measurements were taken at room temperature, which is justified in view of the high Curie temperature of Co$_2$FeSi. Surface sensitive total electron yield spectra (TEY) were taken concurrently with transmission spectra. The transmission signal was collected by a photodiode behind the sample, which measured the x-ray luminescence in the substrate.\cite{Kallmayer07} The beamline resolution was about 0.1\,eV.

Structural characterization was performed by x-ray diffraction in a Philips X'Pert Pro MPD diffractometer with a Cu anode. Bragg-Brentano optics were used for the specular measurements, and an open Eulerian cradle in combination with collimator point-focus optics was applied for off-specular measurements.

Film compositions were determined by x-ray fluorescence spectroscopy, which has been cross-checked with inductively coupled plasma optical emission spectroscopy. The film stoichiometries were thus adjusted to Co$_{50\pm2}$Fe$_{25\pm1}$Si$_{25\pm1}$.

Magnetization measurements have been performed with a vibrating sample magnetometer at 50\,K and 300\,K in magnetic fields up to 3\,T.

\section{Computational approach}\label{Sec:Computational}
The computational work was carried out within the \textsc{Elk} code,\cite{elk} which is based on the full potential linearized augmented plane waves (FLAPW) method. Various approximations to the exchange-correlation functional were tested. Within the generalized gradient approximation (GGA), we chose the Perdew-Burke-Ernzerhof (PBE) functional.\cite{PBE} We performed LDA$+U$ calculations, based on the Perdew-Wang local (spin) density approximation (LDA).\cite{PW92} The fully localized limit (FLL) was adopted, as formulated by Liechtenstein \textit{et al}.\cite{Liechtenstein95} In addition, we performed GGA$+U$ calculations as well. In both cases, we neglect multipole components in the expansion of the Coulomb interaction by using only the U parameter, and setting the intra-atomic exchange parameter $J = 0$. Thus, $U_\mathrm{eff} = U - J = U$, and the Slater integrals $F^{2}$ and $F^{4}$ are zero. In LDA$+U$/GGA$+U$ calculations, one often chooses the $U$ parameter such, that the experimental magnetic moment or band gap is reproduced. 

As another route to obtain an electronic structure with the full magnetic moment, we use the PBE functional and enforce the experimental (or Slater-Pauling) magnetization within a fixed spin moment (FSM) calculation. Thus, we obtain the (variationally unique) PBE band structure with the lowest energy, that results in the desired magnetization.\cite{Schwarz84} This method is usually used to map the dependence of the total energy on the magnetization. Here, we use it to model the unoccupied states of Co$_2$FeSi, such that the minority \textit{d}-states above the gap are shifted above the Fermi energy (see the upper part of Figure \ref{Fig:DOS} for a comparison of GGA and GGA FSM results). We have performed the calculations with a fixed total spin moment of $6\,\mu_\mathrm{B}$\,/\,f.u.

The muffin-tin radii $R_\mathrm{MT}$ were set to $2.2\,$a.u. for all atoms, and the plane-waves cutoff was set to $R_\mathrm{MT} \cdot k_\mathrm{max} = 8.0$. The angular momentum expansion in the muffin-tins was taken to $l_\mathrm{max} = 10$ for both the wave functions and the potential. The maximum $\mathbf{G}$-vector length for the density and potential expansion in the interstitial was set to $|\mathbf{G}|_\mathrm{max} = 14.0$\,a.u.$^{-1}$. $\mathbf{k}$-point meshes of $20 \times 20 \times 20$ points, restricted to the irreducible wedge of the Brillouin zone, were used. Spin-orbit coupling was included by the second-variational method. The 2\textit{p} core states were described as (scalar relativistic) local orbitals, i.e., the $2p_{3/2}$ and $2p_{1/2}$ state have the same basis function. This can give rise to a small error for the $2p_{1/2}$ state, when compared to a fully relativistic description of the core states. Exchange splitting of these states is taken into account. X-ray absorption spectra were computed within the Kubo linear response formalism as lined out in Ref. \onlinecite{Rathgen04}. All theoretical spectra are broadened with a 0.3\,eV wide (FWHM) Lorentzian to account for lifetime broadening.

Experimental lattice parameters of CFS were reported to be 5.64\,\AA{} and 5.66\,\AA{}.\cite{Wurmehl05,Blum09} We used the average value (5.65\,\AA{}) for our calculations.

\section{Results}\label{Sec:Results}

\subsection{Structural characterization}

The samples were epitaxial with a [001] orientation. The out-of-plane lattice parameters are close to the bulk values (Table \ref{Tab:structural_parameters}). The structural ordering is given by the long-range order parameters $S_\mathrm{B2}$ and $S_\mathrm{L2_1}$.\cite{Takamura09} 

L2$_1$ order is observed for both samples, and the order parameters $S_\mathrm{B2}$ are close to 1 for CFS-Cr and CFS-HTD (Table \ref{Tab:structural_parameters}).  The x-ray scattering amplitudes of Co and Fe are very similar, thus we can not detect chemical disorder between Co and Fe. Therefore, $S_\mathrm{B2}$ measures the disorder between Si and Co sites, which is virtually zero. $S_\mathrm{L2_1}$ measures the disorder between Si and Fe sites; we observe a high degree of L2$_1$ order for both samples. While CFS-HTD has approximately full L2$_1$ order, we find a slightly reduced ordering for CFS-Cr, with about 10\,\% of the Fe sites occupied by Si. Note the larger experimental uncertainty for S$_\mathrm{L2_1}$ due to defocussing of the x-ray beam in off-specular measurements.

CFS-HTD has an x-ray reflectometry profile that indicates island growth. The real structure of this film is similar to what was observed by Schneider \textit{et al.}, \cite{Schneider07} i.e., a discontinous film. CFS-Cr has a smooth surface. The rocking curve widths $\Delta \omega$ (reflecting the crystalline orientation distribution) are very different for the two samples, being small for CFS-HTD and rather large for CFS-Cr.

\begin{table}[t]
\caption{Structural properties of the samples: out-of-plane lattice constants, B2 and L2$_1$ order parameters, and rocking curve full width at half-maximum.}
\begin{ruledtabular}
\begin{tabular}{ l l l l l}
Name &  $a$ (\AA{}) & $S_\mathrm{B2}$ & $S_{\mathrm{L2}_1}$ & $\Delta \omega$ \\ \hline
 CFS-Cr &  5.62 & $0.97\pm0.02$ & 0.8$\pm$0.05 & 1.6 \\
 CFS-HTD &  5.65 & $0.98\pm0.02$ &1.0$\pm$0.05  & 0.5 \\
\end{tabular}
\end{ruledtabular}
\label{Tab:structural_parameters}
\end{table}

\subsection{X-ray magnetic circular dichroism}\label{subsec:XMCD}

\begin{figure}[b]
\includegraphics[width=8.6cm]{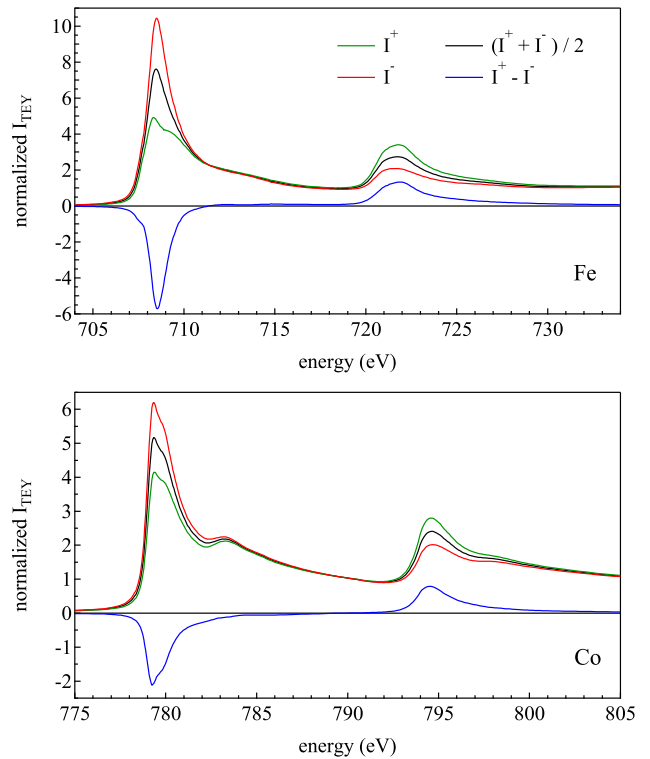}
\caption{\label{Fig:XMCD}X-ray absorption and magnetic circular dichroism spectra of the sample CFS-Cr taken in total electron yield. The absorption spectra are normalized to unity 40eV above the L$_3$ absorption onset.}
\end{figure}

In Fig. \ref{Fig:XMCD} we exemplarily show the x-ray magnetic circular dichroism spectra obtained in total electron yield on sample CFS-Cr. 

\begin{table}[t]
\caption{Element resolved magnetic spin and orbital moments of CFS-Cr obtained from sum rule analysis of the TEY and luminescence spectra. All moments are given in $\mu_\mathrm{B}$.}
\begin{ruledtabular}
\begin{tabular}{ l l l l l }
 	& 	$m_s^\mathrm{TEY}$ 	&	$m_s^\mathrm{Lum}$	&	$m_l$	&	$m_l/m_s$	\\ \hline
 Co &  $1.32\pm0.2$ &$1.38\pm0.2$	& $0.10\pm0.01$ & $0.077$\\
 Fe &  $3.17\pm0.4$ &$3.01\pm0.4$	& $0.09\pm0.01$ & $0.030$ \\
\end{tabular}
\end{ruledtabular}
\label{Tab:moments}
\end{table}

All characteristic features that have been observed by other authors for Co$_2$FeSi are present in our XAS/XMCD spectra as well: the shoulders on the trailing edges of the absorption edges, and the feature in the Co spectra 4\,eV above the absorption onset.\cite{Wurmehl05,Kallmayer11} The latter arises from an \textit{s}-\textit{d} hybrid state of Co and Si. The luminescence spectra (not shown) are in good agreement with the TEY data.

The element specific magnetic moments are extracted from a sum rule analysis of the spectra, assuming the radial $p \rightarrow d$ transition matrix constants $C^\mathrm{Fe} = 6.6$\,eV and $C^\mathrm{Co} = 7.8$\,eV according to Ref. \onlinecite{ScherzPhD}. The spin moment ratio determined from TEY is $m_s^\mathrm{Fe} / m_s^\mathrm{Co} = 2.4\pm0.3$, whereas the ratio from luminescence detection is $m_s^\mathrm{Fe} / m_s^\mathrm{Co} = 2.2\pm0.3$. Both samples have extrapolated low-temperature magnetizations of $6.1\pm0.2\,\mu_\mathrm{B}$\,/\,f.u. (determined by VSM); using this value to scale the XMCD sum rule results, we find element-resolved moments as given in Table \ref{Tab:moments}. The scaling removes experimental uncertainties with respect to the magnetic saturation of the samples.

With the white line sum rule from Ref. \onlinecite{Stoehr95}, we obtain the numbers of \textit{d} holes: $N_h^\mathrm{Co} = 3.0\pm0.4$ and $N_h^\mathrm{Fe} = 4.3\pm0.6$, which are significantly higher than the values for the pure elements (2.4 and 3.45, respectively).\cite{ScherzPhD} The analysis of the sum rules rests on the assumption that the radial $p \rightarrow d$ transition matrix constants are independent of the chemical environment. This might be true within about 10\,\% and gives rise to a corresponding error. Further, there is some uncertainty due to the background subtraction, which we estimate to be of the order of 10\,\% as well.

\begin{figure}[t]
\includegraphics[width=8.6cm]{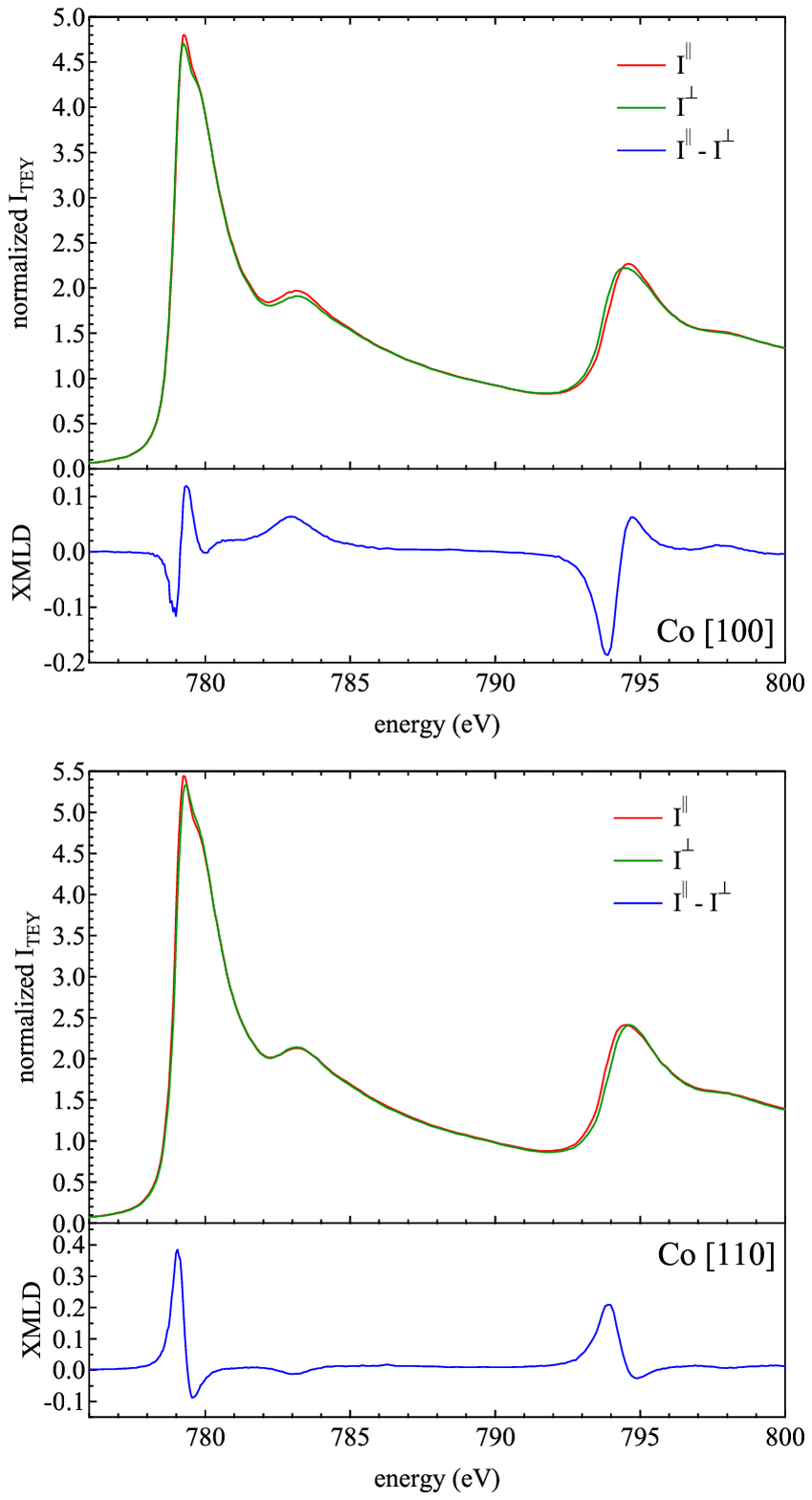}
\caption{\label{Fig:XMLD_Co}Fundamental x-ray magnetic linear dichroism spectra of Co in Co$_2$FeSi, taken in total electron yield. The normalization is the same as in Fig. \ref{Fig:XMCD}.}
\end{figure}

\begin{figure}[t]
\includegraphics[width=8.6cm]{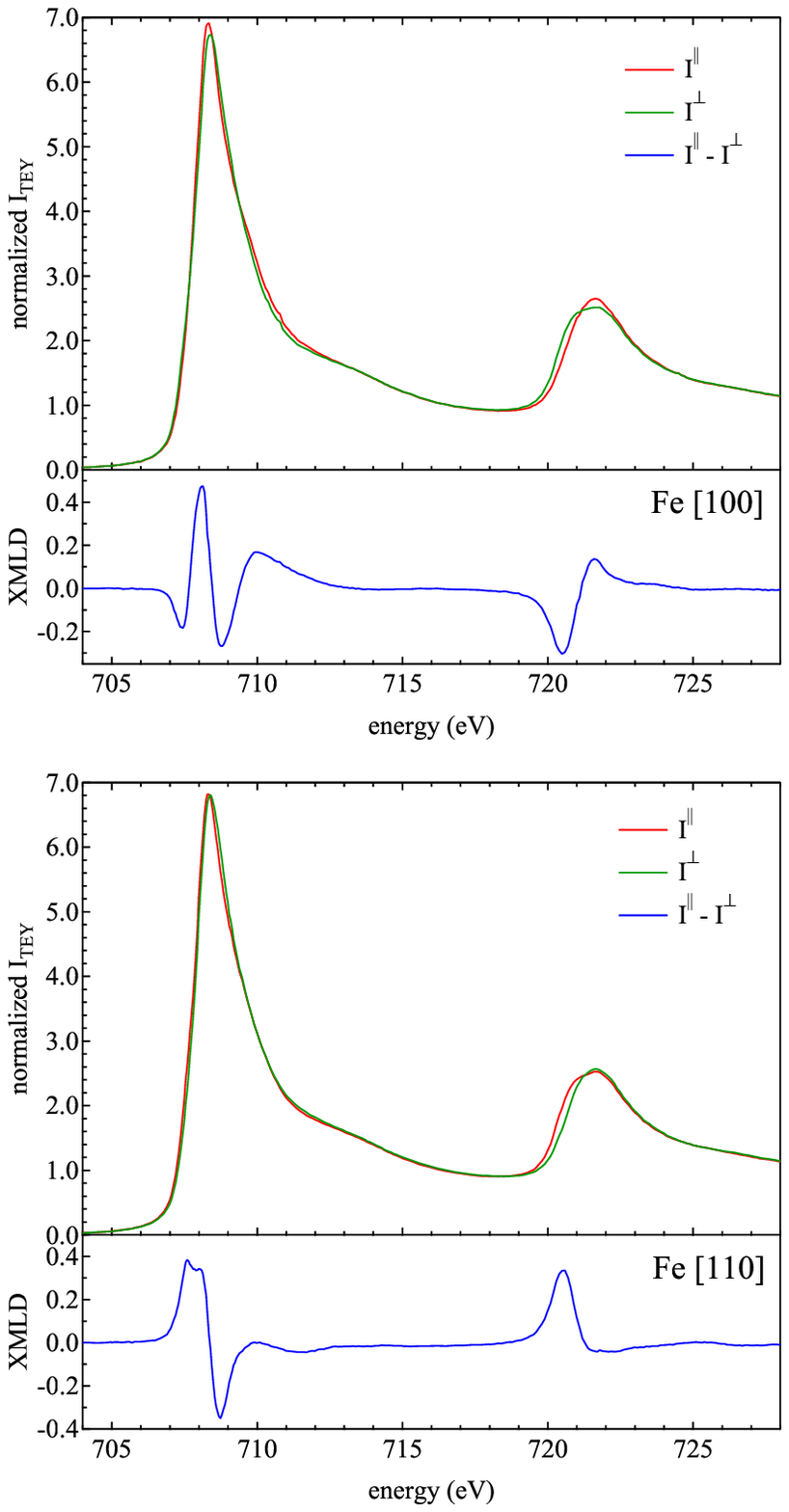}
\caption{\label{Fig:XMLD_Fe}Fundamental x-ray magnetic linear dichroism spectra of Fe in Co$_2$FeSi, taken in total electron yield. The normalization is the same as in Fig. \ref{Fig:XMCD}.}
\end{figure}

\subsection{X-ray magnetic linear dichroism}
The XMLD is generally measured as the difference between two absorption spectra taken with linearly polarized light; one with the polarization vector parallel to the magnetization, and one perpendicular. Expressed with TEY intensities, i.e. $\mathrm{XMLD} = I_\mathrm{TEY}^{||} - I_\mathrm{TEY}^\perp$.  The XMLD has a large crystal anisotropy. It was shown for cubic systems that all XMLD spectra can be described as a linear combination of two fundamental spectra, which are measured along the [100] and [110] high symmetry directions, respectively.\cite{Arenholz06} Earlier reports of XMLD of Heusler compounds include Co$_2$MnSi and Co$_2$MnAl,\cite{Telling08}  Mn$_2$CoGa and Mn$_2$VGa,\cite{Meinert11_1} and Co$_2$TiSn.\cite{Meinert11_2}

The XMLD spectra of CFS-Cr are presented in Figures \ref{Fig:XMLD_Co} and \ref{Fig:XMLD_Fe}. The spectra have pronounced substructures, and the intensities are quite large. For Co, the maximum XMLD intensity is found in the [110] direction at the L$_3$ edge, and amounts to 7.1\,\% with respect to the L$_3$ resonance height. The maximum intensity for Fe is observed at the L$_3$ edge in the [100] direction, and it amounts to 6.9\,\%. For Co, the maximum XMLD is twice as large in the [110] direction than in the [100] direction. In contrast, for Fe the signal strengths are similar in the two fundamental directions.

Comparing the Co XMLD spectra with the available published data of other Heusler compounds, we note that the [110] spectra are very similar in all cases (Co$_2$MnSi, Co$_2$MnAl, Co$_2$TiSn, and Mn$_2$CoGa).\cite{Telling08, Meinert11_1, Meinert11_2} The only exception is Co$_2$MnSi, which has a sharper, and more detailed structure. Telling \textit{et al.} ascribe this (in particular in comparison with Co$_2$MnAl) to the half-metallic character of Co$_2$MnSi and the associated localized character of the \textit{d} electrons.\cite{Telling08} This could hint at the lack of half-metallic character in Co$_2$FeSi.

It has been demonstrated that the XMLD of Heusler compounds depends on the square of the spin moment, modified by the degree of localization of the magnetic moment.\cite{Telling08, Meinert11_1,Meinert11_2, Arenholz10} The maximum Co XMLD intensity along [110] at the L$_3$ edge as a function of the square of the Co spin moment ($4.1\,\%/\mu_\mathrm{B}^2$, using the TEY values for consistency) is weaker than for Co$_2$MnAl ($4.79\,\%/\mu_\mathrm{B}^2$), Co$_2$MnSi ($7.6\,\%/\mu_\mathrm{B}^2$), or Co$_2$TiSn ($5.93\,\%/\mu_\mathrm{B}^2$). The heights of the L$_3$ resonances with respect to the background are comparable among these compounds, so this normalization is reasonable. This hints at not fully localized Co moments, and indicates the lack of half-metallicity as well. Still, Co$_2$FeSi has the largest XMLD observed for Co in a Heusler compound due to its large magnetic moment.

The Co [100] XMLD spectrum of Co$_2$FeSi is different from the spectra of the other compounds mainly by a positive contribution at the maximum of the L$_3$ resonance, which is very weak in the other cases.

XMLD spectra of Fe containing Heusler compounds are not available in the literature. However, our spectra are clearly different from XMLD spectra of pure Fe films grown on GaAs(001).\cite{Nolting10} Notably, the Fe spectra of Co$_2$FeSi are similar to the Mn XMLD spectra of Co$_2$MnSi.\cite{Telling08} The [110] spectra have a double-peak contribution at the low-energy side of the L$_3$ edge and a negative contribution at the high-energy side. The L$_2$ edge is even more similar. For the [100] direction we observe a similar resemblance, but Mn has a small positive contribution at the absorption onset, which is missing for Fe. The maximum XMLD intensity of Fe as defined above ($0.69\,\%/\mu_\mathrm{B}^2$) is close to that of Mn in Co$_2$MnSi ($0.77\,\%/\mu_\mathrm{B}^2$) or Co$_2$MnAl ($0.68\,\%/\mu_\mathrm{B}^2$),\cite{Telling08} and larger than in pure Fe ($0.54\,\%/\mu_\mathrm{B}^2$).\cite{Arenholz10, Nolting10} Fe and Mn occupy the same site of the Heusler structure and experience similar local crystal fields. The larger XMLD intensity of Fe in Co$_2$FeSi compared to pure Fe indicates a more localized character of the magnetic moment. These findings emphasize the high site specificity of the XMLD and its sensitivity to the local crystal field.

\begin{figure}[t]
\includegraphics[width=8.6cm]{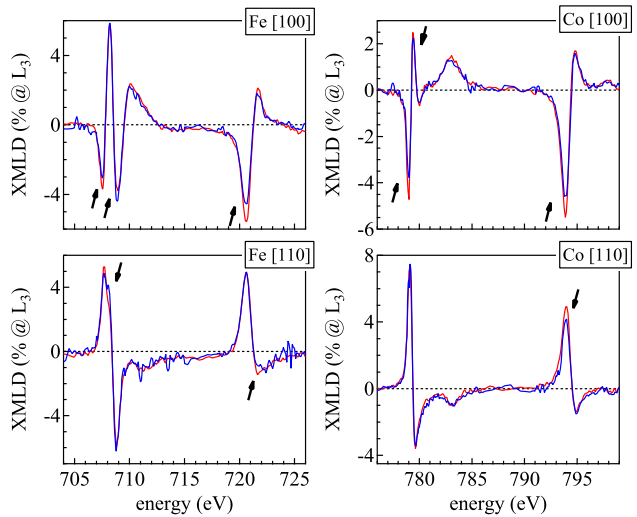}
\caption{X-ray magnetic linear dichroism spectra of the samples CFS-HTD (red traces) and CFS-Cr (blue traces). The spectra were taken in luminescence detection. Subtle differences in the intensities of the various features are visible (see arrows).}
\label{Fig:XMLD_samples}
\end{figure}

In Figure \ref{Fig:XMLD_samples} we compare the XMLD spectra of our samples taken in luminescence detection mode. The intensities are given as percentages of the L$_3$ absorption peak. The main purpose of this comparison is to verify that the XMLD spectra are essentially independent of the deposition conditions and specific details of the films. The various spectra are very similar throughout the samples, and only small differences are seen for the intensities of individual features. These differences may be associated with the slightly lower degree of L2$_1$ ordering in the CFS-Cr sample.

\subsection{First principles calculations}

\begin{figure*}[t]
\includegraphics[scale=1]{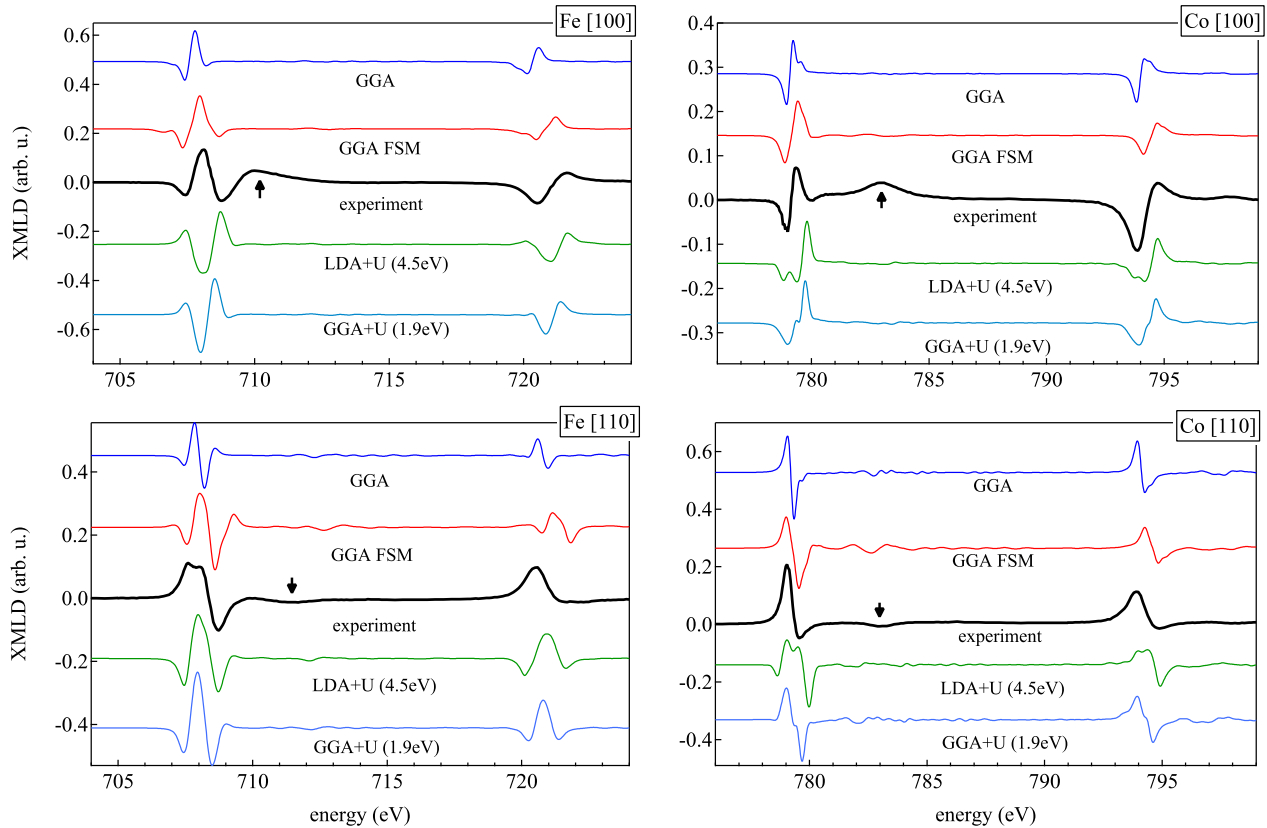}
\caption{\label{Fig:XMLD_theory}X-ray magnetic linear dichroism spectra calculated with different approaches, as discussed in the text. The theoretical spectra are vertically offset against the experimental spectra. They are aligned in energy and scaled to the experimental data.}
\end{figure*}

Four computational approaches are compared with experiment: plain GGA, GGA fixed spin moment (FSM), LDA$+U$ ($U=4.5$\,eV), and GGA$+U$ ($U=1.9$\,eV). For LDA$+U$ we adopt the $U$ value of References \onlinecite{Wurmehl05, Kallmayer11}. In our GGA$+U$ approach, we can use a slightly lower $U$ than in LDA$+U$ to obtain the full magnetic moment. In the latter case, at least 2.5\,eV are required; here, we adopt the smallest value that results in the full magnetic moment. This value was also used in Ref. \onlinecite{Balke06} within a GGA$+U$ approach. The total magnetic moment neglecting spin-orbit coupling with the plain GGA (with the PBE functional) is 5.53\,$\mu_\mathrm{B}$, which is in good agreement with previously reported values.\cite{Kandpal07} Figure \ref{Fig:XMLD_theory} shows the theoretical and experimental spectra for Co and Fe along the [100] and [110] directions.

The general shape of the GGA spectra is quite similar to the experimental ones, but the width of the theoretical curves is considerably too small. The GGA FSM description improves considerably the width and general shape of the spectra, and gives a very good overall resemblance of the experimental data of Co [100] and [110], and Fe [100].

In contrast, both the LDA$+U$ and GGA$+U$ spectra do not describe the experimental data. For Fe [100], the L$_3$ spectra are inverted, and the Co spectra have spurious substructures. Further, the Co spectra are broader than the experimental data. The computed \textit{d} band width is consequently too large.

The Fe [110] spectra are very similar in the GGA, GGA FSM and LDA/GGA$+U$ approaches. All of them have a negative contribution at the absorption onset, which is positive in the experiment. Some features at the high energy side of the L$_3$ edges in the experimental spectra are not captured by any theoretical approach - these are marked with arrows in Fig. \ref{Fig:XMLD_theory}. In the Co spectra, these features are associated with the Co-Si \textit{s}-\textit{d} hybrid state. Since our one-electron theory fails to describe these features, we conclude they are associated with many-body effects.

Electron--hole correlations, in particular the formation of bound excitons (which lead to the so-called 'multiplet' splittings in the absorption spectra), are often observed in the L$_{3,2}$ x-ray absorption spectra of 3\textit{d} transition metal atoms in insulators. However, these effects are virtually absent in the spectra of metals, and reduce to a change in the branching ratio away from the statistical 2:1 in the early 3\textit{d} transition metals. Laskowski and Blaha have shown by solving the Bethe-Salpeter-Equation in an all-electron framework that the dielectric response of the system plays a crucial role for these effects.\cite{Laskowski10} The formation of the splittings is governed by the screened Coulomb interaction of the excited electron with the core-hole. In a (half-)metal, this interaction is exponentially screened out, so that only small changes of the spectra remain. In a recent paper we have demonstrated that this effect is of the order 0.3\,eV for Co in Co$_2$TiSn by means of core-hole calculations in the independent-particle-approximation,\cite{Meinert11_1} whereas the unoccupied \textit{d} band width is of the order 2\,eV here. Thus, due to the metallic screening of the Coulomb interaction it is reasonable to assume that independent-electron calculations can give a good description of the x-ray absorption spectra of Co$_2$FeSi.

From the comparison of the calculations, it becomes clear that GGA FSM provides the best overall matching of the experimental data, and can serve to model the unoccupied band structure of Co$_2$FeSi. Based on the GGA FSM electronic structure we have computed the number of \textit{d} holes of Fe and Co. This has to be done with care in the FLAPW method, because these numbers depend on the choice of the muffin tin radii. Therefore we computed the hole numbers with different choices of muffin tin radii and selected the ones for which the partial magnetic moments saturate - the relevant part of the \textit{d} orbitals is then presumably entirely contained in the muffin-tins. These radii (2.2\,a.u.) were used for all calculations throughout this work (muffin tin radii of 2.28\,a.u. correspond to maximum space filling). The corresponding numbers of \textit{d} holes are $N_h^\mathrm{Co} = 2.9$ and $N_h^\mathrm{Fe} = 4.1$. Both are a bit smaller than the values extracted with the white line sum rule (see Section \ref{subsec:XMCD}). However, they are larger than the theoretical hole numbers of the pure elements, which amount to 2.4 and 3.45, respectively.\cite{ScherzPhD} Our values are close to the numbers of unoccupied \textit{d} states of the free Co and Fe atoms. This atomic-like configuration could be the reason for the failure of the PBE functional to obtain the correct magnetic moment.

The calculated magnetic moments within the GGA FSM approach are $m_s^\mathrm{Co} = 1.49\,\mu_\mathrm{B}$, $m_l^\mathrm{Co} = 0.043\,\mu_\mathrm{B}$ and $m_s^\mathrm{Fe} = 2.98\,\mu_\mathrm{B}$, $m_l^\mathrm{Fe} = 0.043\,\mu_\mathrm{B}$. The interstitial contribution is almost zero. The calculated spin moment ratio is 2, which is smaller than the experimental value. The orbital moments are underestimated by a factor of about 2.3, which is typically observed in DFT calculations without orbital corrections.\cite{Sipr11}

We note that it is still an open question why bulk magnetometry of presumably half-metallic Heusler compounds gives the integer Slater-Pauling values, which are only valid in the complete abscence of valence-band spin-orbit coupling. XMCD does, however, always give small orbital moments, which should increase the total moment.

\section{Discussion}\label{Sec:Discussion}
Kune\v{s} \textit{et al.} proposed a theoretical model of the XMLD, in which the XMLD is given as an energy derivative of differences of the unoccupied $e_g$ and $t_{2g}$ densities of states:
\begin{equation}
\mathrm{XMLD} \propto \Delta_x \frac{\mathrm{d}}{\mathrm{d} E} \left[ \alpha (t_{2g \uparrow} - t_{2g \downarrow}) + \beta (e_{g \uparrow} - e_{g \downarrow}) \right],
\end{equation}
where $\Delta_x$ is the core-level exchange splitting, and $\alpha, \beta$ are constants, which take the values $\alpha = -1$ and $\beta = 2$ for the [100] magnetization direction.\cite{Kunes03,Nolting10} The energy dependence has been omitted for clarity. This model is only valid without spin-orbit coupling, which lifts the degeneracies of the crystal-field split doubly and triply degenerate \textit{d} states. However, this model gives an instructive picture of the local crystal field sensitivity of the XMLD.

Consequently, the large differences observed for the GGA FSM and GGA$+U$ spectra along the [100] direction indicate significant changes in the distribution of the orbital symmetries (irreducible representations) of the \textit{d} states.

\begin{figure}[t]
\includegraphics[width=8.5cm]{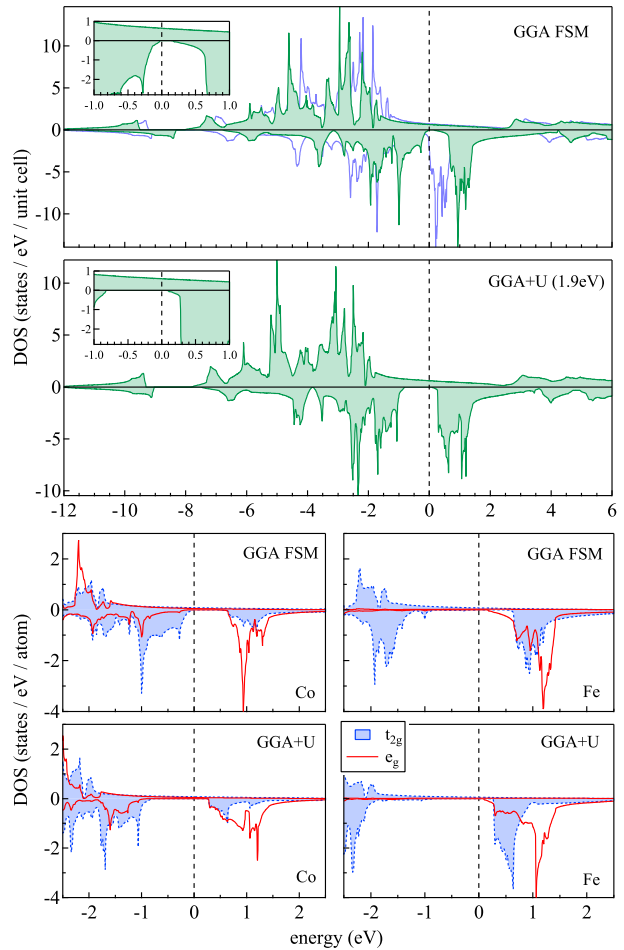}
\caption{\label{Fig:DOS}Top: Total densities of states of Co$_2$FeSi with the GGA fixed spin moment (FSM) and GGA$+U$ (1.9\,eV) approaches. The insets show the vicinity of the Fermi level. The light blue line in the top panel represents the GGA calculation. Bottom: Partial densities of states of Co and Fe in the two computational approaches. The spectra are decomposed into the irreducible representations ($\mathrm{e}_\mathrm{g}$ and $\mathrm{t}_\mathrm{2g}$) of the \textit{d} states. Spin-orbit coupling is neglected.}
\end{figure}

In Figure \ref{Fig:DOS} we compare the total DOS of the GGA FSM calculation and the GGA$+U$ calculation neglecting spin-orbit coupling. The gap widths and positions of the Fermi levels are different, and the width of the unoccupied minority \textit{d} states is larger in the GGA$+U$ calculation. The size of the minority gap in the GGA$+U$ calculation is about $U/2$. The GGA FSM density of states is essentially the same as that of a plain GGA calculation, with the majority and minority states being rigidly shifted against each other. Due to their overall similarity, typical experimental techniques that probe the DOS (photoemission, inverse photoemission, x-ray absorption) are hardly able to distinguish between the models of the electronic structure.

By decomposing the individual \textit{d} projections into their orbital symmetries, we are able to observe differences between the GGA FSM and GGA$+U$ models. This is shown in the bottom part of Figure \ref{Fig:DOS}. The DOS of the two orbital symmetries ($e_g$ and $t_{2g}$) have entirely different structures in the two calculations. The $\mathrm{e}_\mathrm{g}$ and $\mathrm{t}_\mathrm{2g}$ states are more split in GGA$+U$ (i.e., the crystal field is stronger), and the hybridization between Co and Fe $\mathrm{t}_\mathrm{2g}$ states becomes stronger as well. The Co $\mathrm{e}_\mathrm{g}$ states are broader in GGA$+U$, and the Fe $\mathrm{t}_\mathrm{2g}$ states have a larger weight in GGA$+U$ than in GGA FSM. The crystal field splitting is the driving mechanism for the gap creation in the half-metallic Heusler compounds.\cite{Galanakis02} Obviously, the $+U$ operator (in the FLL formulation) enhances the crystal field splitting and forms the half-metallic gap.

It has been shown that the LDA$+U$ formalism can be derived as an approximation to the quasiparticle self-energy within the \textit{GW} method.\cite{Anisimov97, Jiang10} The derivation involves three assumptions: (1) the screened Coulomb interaction (described by $U$) is static; (2) only localized states are subject to the quasiparticle corrections; (3) the many-body exchange interaction between localized and itinerant electrons is neglected. Thus, LDA$+U$ is a kind of on-site Hartee-Fock theory with the effective screening given by the $+U$ parameter. It does provide results that are quite similar to those from the hybrid PBE0 functional.\cite{Nourmohammadi10} The inclusion of Hartree-Fock exchange without the proper spacial and frequency dependence of the screening (i.e., the correlation) allows to obtain the magnetic moment of Co$_2$FeSi correctly, but fails at the correct prediction of the unoccupied states band structure.

The GGA is able to describe the wave functions and unoccupied states reasonably well up to a rigid shift of the latter. Since the GGA can be seen as a static, local, and orbital-independent approximation to the \textit{GW} self-energy, the success of reproducing the XMLD spectra well is based on a cancellation of errors, due to omission of both the non-locality and the frequency dependence in the exchange and correlation contributions. The good balance between exchange and correlation in the LDA (and GGA, which is essentially the same theory) has been confirmed earlier by Kotani and Gr\"uning \textit{et al.} These authors have performed calculations of transition metal and semiconductor band structures with the EXX+RPA approach, which treats exchange and correlation on the same level of many-body perturbation theory.\cite{Kotani98, Gruening06} EXX+RPA reproduced the LDA results well, including magnetic moments and exchange splittings of Fe, Co, and Ni. However, for MnO, where screening is weak, EXX+RPA gave substantial improvement of the band structure.\cite{Kotani98} Furthermore, \textit{GW} calculations on half-metallic CrAs and MnAs by Damewood and Fong\cite{Damewood11} demonstrated that the GGA can predict half-metallic gaps that are very close to the \textit{GW} values, depending on the details of the local fields and the associated screening. 

The GGA FSM band structure has a small gap (about 0.1\,eV wide) in the minority states, and the Fermi energy is located in the gap. Inclusion of spin-orbit coupling closes this small gap. This observation may explain the rather low TMR values of magnetic tunnel junctions with Co$_2$FeSi electrodes, the lack of half-metallic characteristics in the tunneling conductance curves, and the large Gilbert damping.\cite{Kubota09} 

\section{Conclusions}\label{Sec:Conclusions}

Our experimental results show that the features associated with the GGA$+U$ band structure are not present in Co$_2$FeSi. GGA$+U$ predicts an inadequate distribution of the crystal-field split \textit{d} states, which we identified by x-ray magnetic linear dichroism. The $+U$ operator opens a minority gap at the Fermi energy, which necessarily leads to an integer magnetic moment. While exchange and correlation effects beyond the GGA are certainly important for the description of Co$_2$FeSi, the simple $+U$ approach does not produce the correct band structure. The PBE functional is able to produce an unoccupied states band structure for Co$_2$FeSi which is approximately correct up to a rigid shift of the minority \textit{d} states. 

Co$_2$FeSi is a complex material, being neither close to the electron gas (the GGA fails to find the correct ground state magnetic moment), nor being strongly correlated (GGA$+U$ fails at the description of spectral properties). That calls for further refinement of the approximations to the density functional. More flexible functional forms, such as meta-GGA functionals (adding information on the kinetic energy density), may perform better in this respect. Co$_2$FeSi is therefore an interesting test case for new density functionals. Ultimately, many-body approaches, such as EXX+RPA,\cite{Kotani98} the \textit{GW} approximation (possibly including some kind of self-consistency), or DMFT\cite{Minar11,Biermann03} might be required to obtain a consistent theoretical picture of Co$_2$FeSi.

\begin{acknowledgments}
The authors gratefully acknowledge financial support from Bundesministerium f\"ur Bildung und Forschung (BMBF) and Deutsche Forschungsgemeinschaft (DFG, Grant RE 1052/22). They thank for the opportunity to work at the Advanced Light Source, Berkeley, CA, USA, which is supported by the Director, Office of Science, Office of Basic Energy Sciences, of the US Department of Energy under Contract No DE-AC02-05CH11231. Special thanks go to the developers of the \textsc{Elk} code.
\end{acknowledgments}

\end{document}